# Machine Translation in the Wild: User Reaction to Xiaohongshu's Built-In Translation Feature

**Sui He**
School of Culture and Communication
Swansea University
United Kingdom
sui.he@swansea.ac.uk

## Abstract

This paper examines user reactions to the launch of the machine translation (MT) feature on Xiaohongshu, a Chinese social media and e-commerce platform, in January 2025. Drawing on a dataset of 6,723 comments collected from 11 official posts promoting the translation function, this paper combines sentiment analysis with thematic analysis to investigate how users perceived and experimented with the function. Results show that reactions were generally positive, although concerns regarding functionality, accessibility, and translation accuracy were also expressed. In addition, users actively tested the function with inputs that are atypical for everyday online communication, including stand-alone words and phrases, abbreviations, internet slang, and symbolic or encoded forms. Successful decoding of these texts elicited positive responses, while testing of more conventional language remained fairly limited. This could lead to uncritical acceptance of MT outputs by users, highlighting the importance of closer collaboration among computer scientists, translation scholars, and platform designers to improve MT performance and promote informed user engagement in real-world scenarios.

## 1 Introduction

Xiaohongshu ('little red book'), also known as 'RED' or 'rednote', is a Chinese social media and e-commerce platform founded in June 2013 (Xiaohongshu, 2025). The platform combines user-generated content with integrated e-commerce functions, allowing users to share experiences, recommendations, and consumption-related information. Today, Xiaohongshu operates both as a mobile app and as a website, hosting a wide range of lifestyle-related content including fashion, cosmetics, personal care, food, travel, entertainment, reading, fitness, and parenting, etc. Posts on the platform collectively generate over seven billion impressions per day, illustrating the scale of user engagement within the community (ibid.).

Looking at Xiaohongshu's development over the years, a recurring theme is its orientation towards international markets. The original app was known as 'Hong Kong Shopping Guide', targeting Chinese tourists seeking shopping recommendations outside mainland China (Reuters, 2025). In December 2013, six months after the company's founding, Xiaohongshu launched a community dedicated to sharing overseas shopping experiences – an initiative that the company itself highlighted as one of the key milestones in its early development (Xiaohongshu, 2025). Subsequent initiatives further reinforced this international vision. In May 2017, the company introduced ReDelivery, an international logistics service designed to facilitate cross-border e-commerce. Later that year in August, Xiaohongshu established its Global Technology Headquarters in Wuhan, China. In an interview, one of the founders emphasised their global vision and their ambition to position the platform as a space for sharing lifestyle experiences for Chinese and overseas users alike (Bloomberg, 2018).

As the platform gradually expanded its international reach, issues of cross-lingual communication became increasingly relevant. While Xiao-hongshu had long been oriented primarily towards Chinese-speaking users, its global aspirations and cross-border commerce infrastructure created conditions in which multilingual interaction could become more prominent. This dynamic saw a turning point in January 2025, when the platform experienced a sudden influx of international users.

Following renewed discussions about a potential ban on TikTok in the United States in the beginning of 2025, Xiaohongshu attracted a large number of international users. The literature shows that more than three million English-speaking users migrated to the platform (Feng et al., 2026), and the influx of these users propelled Xiaohongshu to become the most downloaded application on Apple's US App Store on 13 January 2025 (Xiao and Zhang, 2025). Many of these users identified themselves as 'TikTok refugees', and discussions surrounding this phenomenon on Xiaohongshu accumulated over 2.45 billion views within days (Yuan et al., 2025).

The surge of this multilingual community revealed the limitations that language barriers imposed on cross-cultural communication on the platform. In response to these challenges, only a couple of days later since the influx, Xiaohongshu promptly introduced a machine translation (MT) function powered by large language models (LLMs) on 18 January 2025, announced by an official post titled bilingually as '小红书翻译功能来啦! Translation is coming'. As any fluent bilingual speaker of Chinese and English would notice, let alone professional translators, this title is far from a good translation: the Chinese title, meaning 'Xiaohongshu's translation function is here', is clear and informative, whereas the English text can be ambiguous and confusing when read on its own. Following this post, a series of official posts were published over the subsequent week, introducing and promoting the MT feature across different use scenarios. These posts collectively formed an informational package, making users aware of the feature's existence and use cases, such as translating posts, comments, and chats.

Building on this launch context, this study investigates user responses to the launch of Xiaohongshu's built-in MT function in its launch context. By analysing users' reactions, comments, and the texts they used to try out the translation feature, the study examines how the new functionality was perceived and experimented by users within the community while also reflecting on the broader lessons that can be drawn from this pioneering attempt to move LLM-powered MT from laboratory settings into mass-user communication. Specifically, it addresses the following research questions:

1. How did users perceive the MT function deployed in the four use cases in Xiaohongshu (i.e., translating posts and comments, translating direct messages in chats, creating bilingual multimodal post content, and generating bilingual subtitles)?

2. When trying out the MT function, what types of texts did users choose?

## 2 Literature Review

MT has become an essential technological feature for facilitating multilingual interaction on social media platforms such as Instagram, TikTok, and X (formerly, Twitter). On these platforms, the MT systems are expected not only to translate textual content, but also support multilingual sense-making and interaction that allow users from different linguistic backgrounds to participate in shared communicative spaces (Lim et al., 2018). However, idiosyncratic features of social media content present particular challenges for MT systems (Carrera et al., 2009; Vieira and Al Sharou, 2025) and the increasing globalisation trend intensifies the demand for effective MT-assisted cross-lingual communication (Gao et al., 2024). Despite the increasing availability of built-in translation functions on social media platforms, the effectiveness and social implications of these systems remain contested (Gupta et al., 2023).

Compared with other social media platforms, limited information is available regarding the translation feature of Xiaohongshu with their publicly accessible documentation provides minimal details. The only mentioning is in the User Privacy Policy[1]: 'When you or other users use the Translation feature, we will provide the content that needs to be translated to the translation service provider. We will only share the content that needs translation and will not disclose usernames or any other personally identifiable information'.

In the literature, several technical papers have presented models that are reportedly designed for or tested on Xiaohongshu data. For instance, Guo et al. (2025) introduce RedTrans, a 72-billion-parameter large language model (LLM) , which is described as having been deployed in a real-world production environment associated with Xiaohongshu. Similarly, Zhao et al. (2025b) present RedOne, a domain-specific LLM designed to support multiple social media related tasks, where translation is included in the list. In a subsequent iteration (Zhao et al., 2025a), the authors introduce RedOne 2.0, which has been reported to be deployed on a large-scale social networking platform with millions of users to generate personalised post titles in real time, implying potential integration within Xiaohongshu's infrastructure. Other research explores multimodal translation tasks related to Xiaohongshu content. For example, Feng et al. (2025) propose MT$^3$, a text-image MT model designed to translate text embedded within images between English and Chinese.

[1] https://agree.xiaohongshu.com/h5/terms/ZXXY20251205003/-1

From a social science and human-computer interaction perspective, research examining the translation function on Xiaohongshu remains scarce. As the feature is relatively new, most existing studies mention translation only as a secondary element rather than a primary research focus.

One strand of literature emphasises the transformative potential of LLMs for global communication. For example, Yan et al. (2025) argue that LLMs enable new forms of international interaction by dismantling barriers of language and culture. They use Xiaohongshu as an example, stating that the platform's LLM-driven translation tools allow users from diverse linguistic backgrounds to communicate 'seamlessly' in real time. Notably, such claims found in literature often rely on technological optimism and do not examine translation accuracy or user experiences in practice.

Other studies take a more critical perspective. Yuan et al. (2025) note that the built-in translation feature on Xiaohongshu primarily supports dialogue comprehension but struggles with more complex linguistic phenomena such as humour, memes, or metaphorical expressions. These elements are particularly prevalent in social media discourse and therefore represent important limitations for automated translation systems.

More recent research further highlights the cultural dimensions of translation on the platform. Feng et al. (2026) analyse how Chinese users translated the usernames of so-called 'TikTok refugees' who migrated to Xiaohongshu. The authors argue that LLMs frequently fail to capture these layered cultural references and therefore cannot fully reproduce the dynamics of cross-cultural communication on social media. They consequently call for further research exploring how users negotiate cultural authenticity, linguistic accessibility, and platform affordances in multilingual digital environments.

Despite these emerging insights, a significant gap is evident in the literature: While technical studies focus on model development and performance, and social science research occasionally references translation as a supporting feature, there is little empirical investigation into how the translation function on Xiaohongshu was experienced by users when it was first launched, and the perceptions, interpretations, and practices of users engaging with the platform's translation tools remain largely unexplored. Addressing this gap is especially important given the platform's growing international user base and the increasing role of MT in shaping intercultural and interlingual communication within social media ecosystems.

## 3 Methodology

### 3.1 Data Collection

Data collection was conducted shortly after the launch of Xiaohongshu's built-in translation function, which became available on the evening of 18 January 2025 (GMT+8). To capture early-stage user interactions while minimising potential data loss due to deletions of posts and comments and potential account closures and restrictions, the dataset was collected after a 10-day observation window on 29 January 2025 (GMT).

The process comprises two stages. First, all official posts announcing or promoting the translation function were manually identified on the platform. This resulted in a set of 11 posts (See Appendix A for an overview). These posts feature four primary use cases promoted by the platform: (1) translating texts in posts and comments (P01–P04, 'post content and comments' in tables and figures); (2) translating texts in direct messages in chats (P05–P07, 'direct messages'); (3) creating bilingual text-embedded post images based on monolingual textual inputs (P08–P09, 'bilingual posting'); and (4) generating English–Chinese bilingual subtitles for videos (P10-P11, 'bilingual subtitling'). Figure 1 presents an overview of the key engagement indicators of the 11 posts.

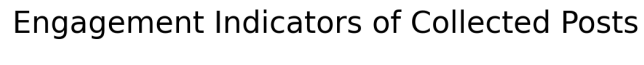

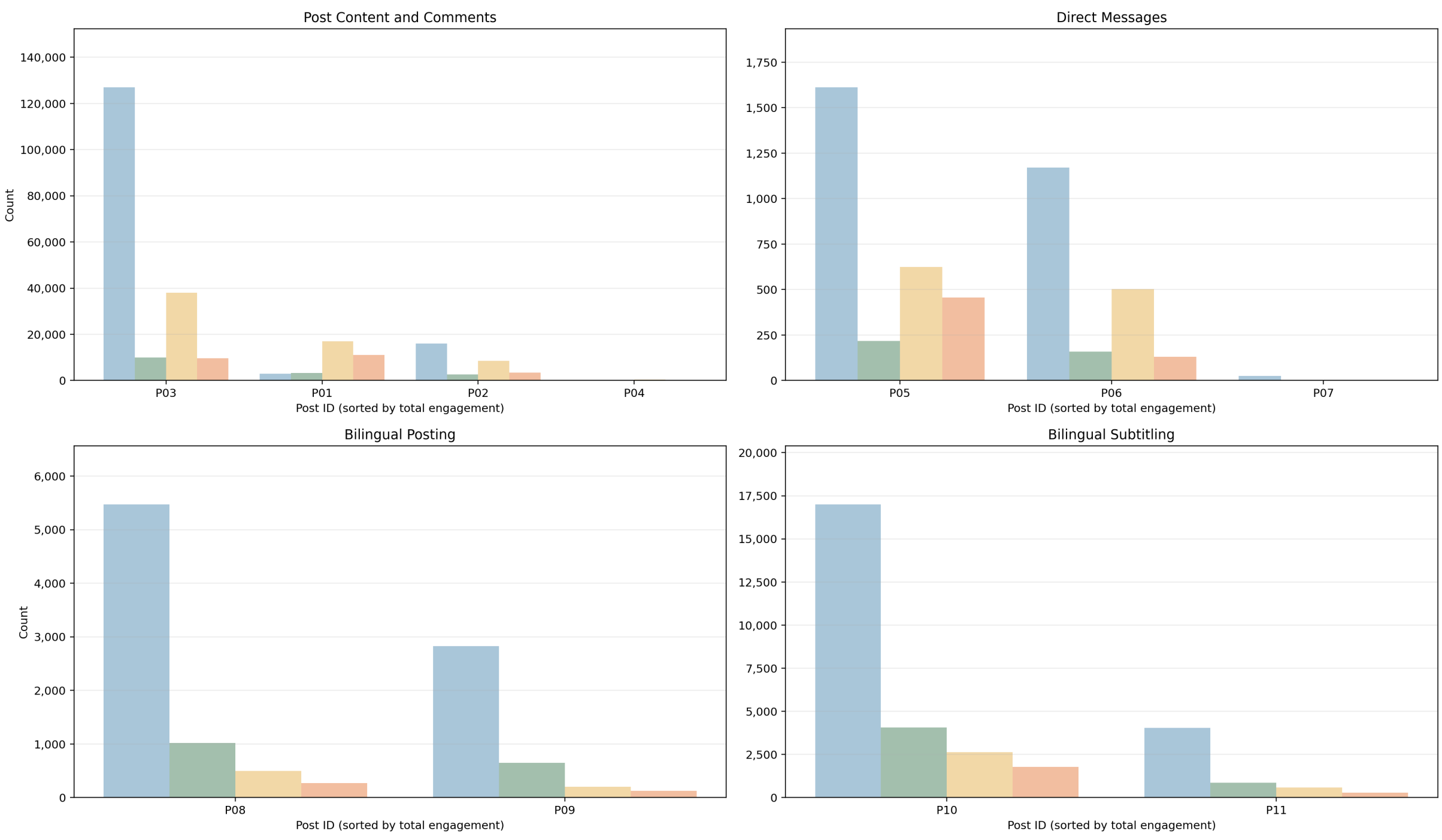

**Figure 1:** Engagement Indicators for Posts

Second, comments associated with these posts were retrieved using the open-source web crawling tool `MediaCrawler`.[2] For each of the 11 identified posts, the crawler was used to collect the top 300 first-level comments (as ranked by the platform's default sorting and visible to generic users within the application interface), along with their sub-comments, which typically contained reflections on the translation of segments in the first-level comments or additional testing segments. This approach ensured that the dataset reflected the most visible and highly engaged user discussions on the newly introduced feature.

In total, 7,154 comments were collected. All data analysed in this project were publicly accessible at the time of collection. No attempts were made to access restricted or private content. During data processing, user identifiers were removed or anonymised, and the dataset was used solely for aggregated analysis.

## 3.2 Analytical Framework

### 3.2.1 Data Cleaning

Prior to analysis, the dataset was pre-processed to remove elements irrelevant to the defined tasks. User identifiers were removed to ensure anonymity and compliance with data protection practices. Comments containing only user mentions (e.g., @USERNAME), formatting commands (e.g., `\n`), or empty content were excluded. After data cleaning, the dataset contained 6,723 comments.

### 3.2.2 Sentiment Analysis

Sentiment analysis, also known as opinion mining, is a computational approach used to identify and extract subjective information, including emotions, attitudes, and evaluations, from textual data (Mughal et al., 2024; Zucco et al., 2020). Social media discourse presents unique challenges for sentiment analysis: informal language, sarcasm, and rapidly evolving linguistic conventions can complicate automated sentiment detection, as the intended meaning may differ from the literal wording of a comment (Wankhade et al., 2022).

Early sentiment analysis research primarily focused on document-level or sentence-level classification. More recent developments have expanded toward aspect-based sentiment analysis (ABSA), which captures opinions directed toward specific targets or features within a text (Wu et al., 2025). Such methods are particularly valuable for analysing large volumes of user-generated con-tent on social media platforms, where individuals continuously share experiences and evaluations in highly dynamic online environments. Additionally, recent advances in LLMs have demonstrated substantial potential for improving sentiment analysis performance in complex linguistic environments (Zhang et al., 2024).

[2]https://github.com/NanmiCoder/MediaCrawler

To address these challenges and account for the complex linguistic and multimodal nature of the dataset, this study adopts a triangulated LLM-plus-human approach using two flagship models (`GPT-5 mini`[3] and `DeepSeek-V3.2`[4]) supplemented by human annotation of randomly selected samples following an aspect-based approach. Each comment was independently classified by the two models, and their outputs were compared to derive consensus results (n = 4519; 67.22%). To assess the reliability of the automated classifications, a randomly selected 10% subset (n = 452) of the consensus-labelled data, distributed across the four use cases, was manually annotated by a human coder. Agreement between the human annotations and the model consensus results was evaluated using Cohen's kappa ($\kappa = .81$), indicating a high level level of agreement.[5] This mixed approach combines the scalability of automated analysis with human validation, helping to mitigate potential model bias and improve the robustness of the analytical results. Table 1 shows an overview of the collected and analysed data across the four use cases.

| Use Case | Raw | Cleaned | Consensus |
|---|---|---|---|
| Post Content and Comments | 5,834 | 5,418 | 3,617 |
| Direct Messages | 312 | 307 | 216 |
| Bilingual Posting | 90 | 90 | 67 |
| Bilingual Subtitling | 918 | 908 | 619 |
| Total | 7,154 | 6,723 | 4,519 |

**Table 1**. Data Overview

### 3.2.3 Prompt Design

To classify user comments, a structured prompt was designed and applied consistently across the two LLMs, following the ABSA approach. The prompt instructed the model to perform a hierarchical classification task separating comment type from sentiment polarity, with contextual information provided and aspect specified for the task (full prompt in Appendix B).

First, each comment was categorised according to its communicative intent: Opinion (expressing evaluation of the translation feature), Test (inputs primarily used to try the translation function), or Other (irrelevant discussion, platform requests, or unrelated interaction). Sentiment polarity (positive, neutral, negative) was then assigned only to comments classified as Opinion. To improve interpretive consistency, the prompt included explicit decision rules clarifying category boundaries and prioritisation when multiple signals appeared. Given that comments on the platform frequently include platform-specific emoji tokens (e.g., '[点赞R]' *press like* or '[笑哭R]' *tears of joy*), the prompt also instructed the models to treat such tokens as potential indicators of sentiment.

In addition, the prompt incorporated a small set of manually selected few-shot examples representing each classification category. Few-shot prompting has been shown to improve model performance in complex tasks such as ABSA, as it provides structured guidance for LLMs that helps stabilise model outputs and regulate response format (Zhang et al., 2024). These examples were therefore included to reduce classification ambiguity and improve consistency across model outputs. To facilitate automated parsing and ensure comparability across model outputs, the models were required to return results in a constrained JSON format.

During inference, model parameters were standardised as far as possible in the two models. The temperature parameter was set to 0 for `DeepSeek-V3.2` (this was not supported by `GPT-5 mini`) to minimise stochastic variation, while `top_p` in each model was set to 1 and batch sizes were kept consistent during processing. Prompt template was tested and polished before being applied to the workflow.

### 3.2.4 Thematic Analysis

To complement the computational sentiment analysis, qualitative thematic analysis (Braun and Clarke, 2008) following an inductive approach was conducted using NVivo 15. Comments labelled as Opinion were examined to identify recurring themes in users' evaluations of the four use cases of the translation feature, including aspects perceived positively, neutrally, and negatively. For this analysis, coded samples within each sentiment cate-gory were randomly selected in proportion to the size of each use case, with 10% of the comments from each case included in the qualitative sample. In total, 304 comments were analysed from the set of consensus data labelled as Opinion comments (n = 3,045).

[3]https://developers.openai.com/api/docs/models/gpt-5-mini
[4]https://huggingface.co/deepseek-ai/DeepSeek-V3.2
[5]Human annotation revealed occasional misclassifications in the consensus-labelled data, particularly in comments containing only 'hahahaha' or slang expressions involving swearing words used to intensify excitement. As detailed analysis of these patterns lies beyond the scope of the present study, they are not examined further here.

In addition, comments labelled as Test were analysed qualitatively to explore how users interacted with the translation feature in practice. A random sample comprising 20% (n = 136) of the test comments (n = 679) was selected for thematic analysis. This analysis aimed to identify patterns in user experimentation with the feature and the types of linguistic inputs used during testing.

## 4 Results

### 4.1 Sentiments

Results from the sentiment analysis indicate that user responses to the translation feature were predominantly positive across the four use cases: among the Opinion-labelled comments (n = 3,045), 78% expressed positive sentiment (n = 2,364), while 4% were neutral (n = 133) and 18% were negative (n = 548). For sentiment distribution across the four use cases, Figure 2 provides an overview.

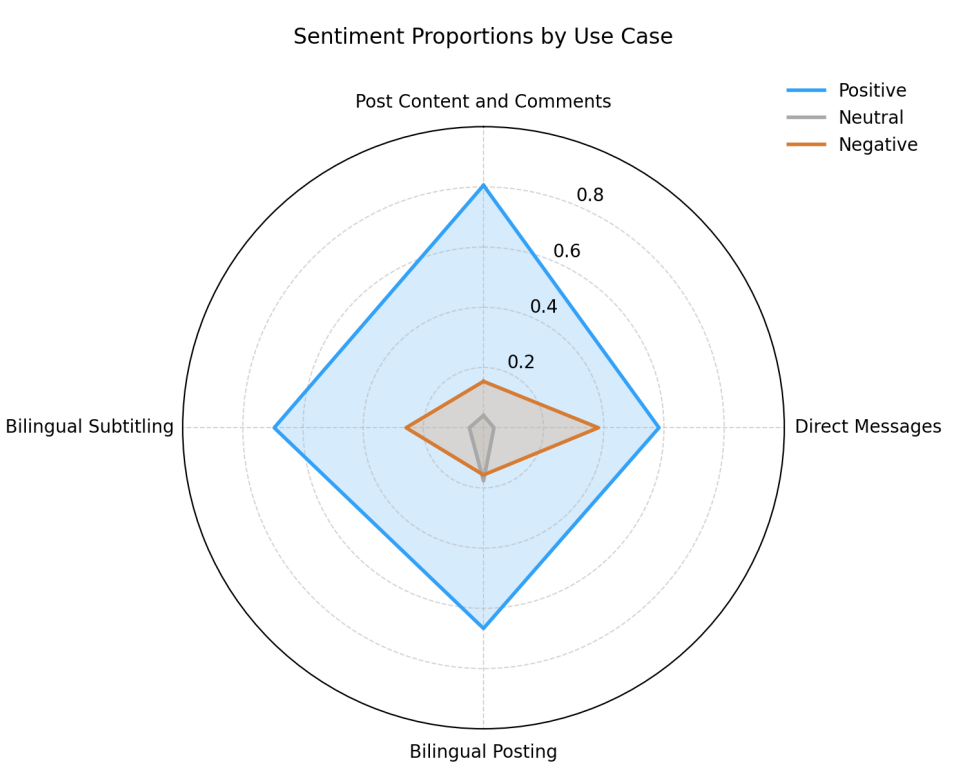

**Figure 2:** Sentiment Proportion by Use Case

As shown in Figure 2, sentiment distributions varied across the four use cases. Comments associated with post content and comment translation constituted the largest share of the dataset (n = 2,403), with the highest proportion of positive sentiment (n = 1,935; 80.5%) across the four use cases and the lowest of negative (n = 370; 15.4%). For direct message translation (n = 144), as it hit the lowest for both positive (n = 84; 58.3%) and neutral (n = 5; 3.5%), the proportion of negative comments (n = 55; 38.2%) was unsurprisingly the highest.For bilingual posting (n = 51), although it showed a generally positive distribution (n = 34; 66.7%), it had the highest proportion of neutral comments (n = 9; 17.6%) and the negative responses (n = 8; 15.7%) were at the lower range. Bilingual subtitling (n = 447) showed a more balanced distribution, where the proportion of positive (n = 311) and negative (n = 115) comments reached 69.6% and 25.7% with neutral staying at 4.7% (n = 21).

To further examine how users evaluated the translation feature, the results from the thematic analysis of selected samples identified key themes reflecting specific aspects of user feedback, as shown in Figure 3.

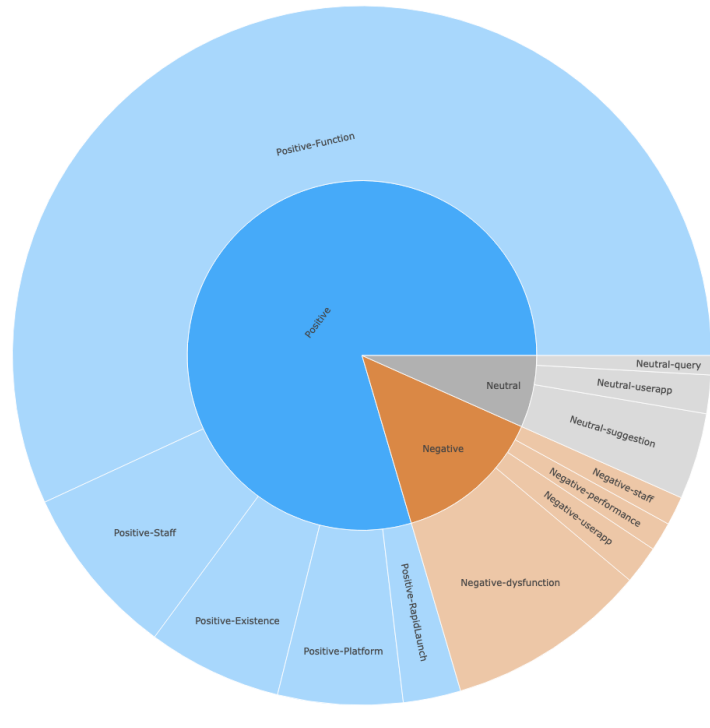

**Figure 3:** Thematic Analysis of Sentiments

For positive comments, the most frequently occurring theme was on the translation function, where users commented favourably on the translation performance, especially for translating certain texts such as abbreviations and internet slangs (e.g., 网络热词也可以翻译 *it can translate popular internet expressions as well*), or the usefulness of the feature in general (e.g., 好厉害的翻译 *outstanding translation*). Other positive themes included appreciation of the existence of the feature itself (e.g., 终于有翻译功能了 *a translation function at last*), references to the platform when praising the feature (e.g., 小红书超厉害 *Xiaohongshu is super*), and expressions of gratitude towards the developers or technical staff responsible for implementing the function (e.g., 给程序员加鸡腿辛苦啦 *give more chicken legs to the programmers; thanks for the hard-work*). A smaller number of comments also highlighted the speed with which the feature was introduced.

Neutral comments largely consisted of practical inquiries or suggestions related to the feature (e.g., 能不能做一个视频字幕带翻译功能呀 *can you enable the translation function for subtitles in videos*). Some inquiries of this kind were promptly addresses by the platform, as can be seen in the positive comments above. In addition, some users asked questions about how the translation function works (e.g., 这是什么语言什么翻译原理 *What language is this and what is the translation mechanism of this*), while others proposed improvements (e.g., 建议再优化一下翻译功能文字里面带表情图的没有翻译选项 *The translation feature could be improved. Text embedded in images with emojis don't have the translation option*). A small number of comments involved users sharing information about their device system or application version when explaining why they could not access the feature.

Negative themes primarily concerned technical issues or limitations of the feature. The most common negative theme was dysfunction, where users reported problems with the translation function not operating as expected (e.g., 我为啥看不到有翻译啊 *why can't I see the translation*). Additional negative comments referred to translation quality or performance issues (e.g., 小红书怎么被翻译成了红笔记 *how come Xiaohongshu was translated as Hongbiji*), incompatibility with certain devices or application versions, and occasional remarks concerning staff welfare or working conditions.

### 4.2 Test Segments

The analysis on test segments posted by users shows several recurring categories, as shown in Figure 4. The most common category consisted of natural language inputs, including texts in multiple languages such as English, Chinese, Cantonese, Chinese (Min), Burmese, Japanese, Russian, Malay, and Latin. In addition, users frequently experimented with language varieties and creoles, including Singlish (Singapore English) and Chinglish (Chinese English). In the English and Chinese segments, most entries consist of greeting (e.g., 外国朋友你们好 *hello foreign friends*), isolated lexical items (e.g., abandon – which often appears among the first words in English vocabulary textbooks), or proper nouns (e.g., KFC).

A second group of test inputs involved artificial or styled language formats, echoing the content of one of the official posts P04 that featured a string of coded text in its title: 请翻译：□●□○○●▲ (*Please translate:* □●□○○●▲). This group included coding schemes such as American Standard Code for Information Interchange (ASCII) (e.g., '73 76 79 86 69 89 79 85' was translated as 'I LOVE YOU'), JavaScript, Braille, and Morse code, as well as kaomoji (e.g., '/□□••' was translated as 猫猫叹气, a meow sigh) and stylised codes (e.g., '●○●●○●●○●' was translated as 星星之火可以燎原, a single spark can start a prairie fire – a well-known saying in Chinese).

Another prominent category consisted of pinyin-based[6] abbreviations commonly used in fandom discourse (e.g., 'jntm' for 'ji(zhiyin)nitaimei', name of a song), proper nouns (e.g., 'cxk' for 'Cai Xu Kun', name of an idol), and slang expressions (e.g., 'xswl' for 'xiao si wo le', meaning *I'm dying laughing*). In addition, users also tested non-pinyin-based internet slang (e.g., 'cpdd', *Looking for a CP (couple in games), DM me*), which often relies on culturally specific expressions or unconventional character combinations. Some users experimented with classical Chinese prose, featuring stylistically and historically distinct texts that impose significant challenges for translation even by translators. A range of other inputs were observed, including strings composed primarily of emojis, numbers, and punctuation.

A small subset of comments contained prompt-style inputs, where users appeared to experiment with instructions or structured prompts rather than conventional text. This behaviour is likely related to the fact that the translation function was powered by LLM at the time of launch, enabling users to temporarily interact with the system in a prompt-like manner before this behaviour was later restricted by the platform.

## 5 Discussion

### 5.1 User Feedback to the New Function

The built-in MT function on Xiaohongshu represents a pioneering effort to deploy LLM-based MT on a mass-user platform outside the lab environment, providing valuable insights into both successes and shortcomings. In terms of its basic configuration, the built-in function was activated based on users' system language settings. For users whose system language was set to a language other than Chinese, the translation function was automatically enabled for Chinese content on the platform. Conversely, for users with Chinese as their system language, the function became available when non-Chinese content was shown. Yet, this information was not communicated clearly in the official posts

[6]Pinyin is the official romanisation system of Chinese characters.

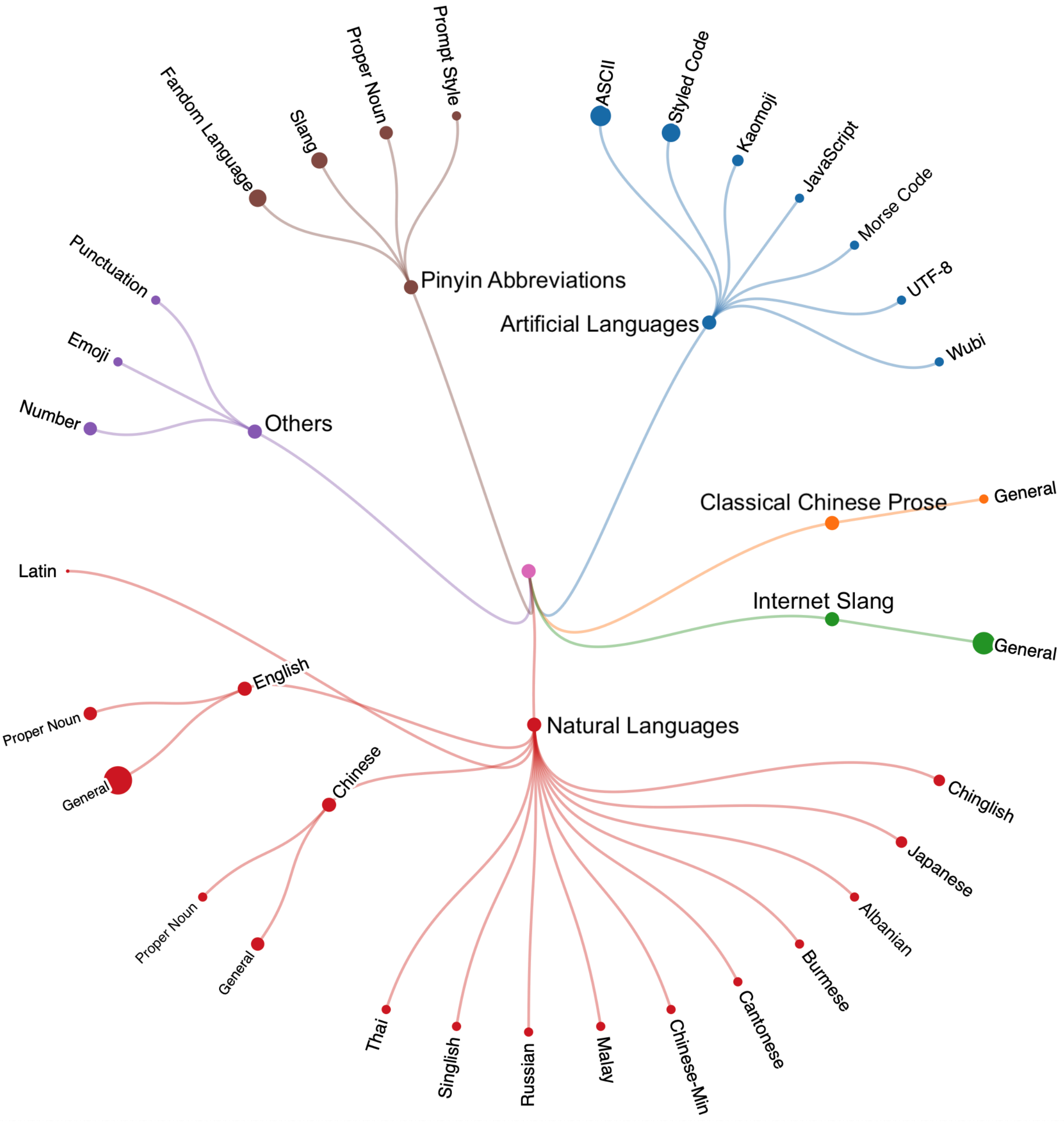

**Figure 4:** Thematic Analysis of Test Segments

– only 'click the translate tab' and 'update to the newest version' were mentioned, and when it was tested in multilingual settings, problems started to emerged, with users complaining not seeing translations they expected or not seeing the translation option at all.

User responses further suggest that perceptions of the translation feature differed across use cases, indicating perceptions of MT functionality were shaped by the context in which it was used. Among the four use cases examined, translating posts and comments was generally perceived most positively. By contrast, multimodal applications including bilingual posting and subtitling received comparatively less favourable reactions. The comparatively weaker reception of multimodal features reflects how contextualised interface design shaped users' evaluation of the MT function. Text-based translation of posts and comments was supported by clear signposting of a clickable 'translate' option next to translatable content, which fits users' intuitive expectations of how the function should work. In contrast, accessing translation in chats requires users to long-press the text before the 'translate' option appears. Both interaction mechanism were clearly presented in the official posts but the key difference was with the interface design. These multimodal use cases seem to have introduced additional layers of complexity, compounded by the absence of efficient onboarding information for users, as reflected in the negative comments on usability and accessibility.

The platform's subsequent consolidation of language settings and multimodal functions into a broader 'content translation' function embedded in the application settings suggests that early user feedback may have informed later design adjustments in a retrospective manner. Reflecting on this trajectory of updates, it becomes evident that launching an MT tool requires careful planning that extends beyond technical model performance. Early deployment should incorporate a comprehensive launch strategy that accounts for user experience, ideally validated through pilot studies, rather than moving directly from a lab setting with limited stakeholders to full-scale public release. While practical pressures such as time constraints and business priorities inevitably influence deployment decisions, considerations of quality and usability should remain central when releasing an MT tool to a mass audience. Such an approach can be guided by emerging research-based, evidence-driven ethical frameworks (Briva-Iglesias and O'Brien, 2026), ensuring reliability, accessibility, and trustworthiness for all users.

### 5.2 Evaluation of Translation Quality

Another key theme emerged from the type of source texts that users chose when testing the feature. A noticeable proportion of comments focused on the system's ability to decode unconventional inputs in textually and/or visually encoded forms. [7] The results show that the successful translations of these inputs often generated enthusiastic reactions, which users interpreted as evidence of the MT system's advanced performance. At the same time, relatively few tests involved complex sentences or extended exchanges in everyday languages. As a result, positive impressions of translation performance were frequently formed on the basis of these atypical or simplified inputs rather than on sustained multilingual communication.

Source texts that users experimented with for the MT feature were very different from the source texts proposed by Xiaohongshu users for language learning purposes as report in Chen (2026). When testing artificial codes to encode texts or symbolic forms to deliver intended meaning, users typically know the expected output and can therefore verify the correctness of the translation directly. In contrast, when translating everyday languages that they understand very little, users often lack the linguistic knowledge or even confidence required to evaluate the accuracy of the output, not to mention the cultural knowledge to evaluate its appropriateness. This is in line with Qiu and Pym (2025), where MT errors evidenced language learners' lack of self-confidence by trusting the (mis-)translated subtitles more than themselves. For users' engagement with MT outputs examined in the current study, it is more likely to be the case where users were not confident in their second language at all, if there is any. The absence of linguistic knowledge as such can potentially lead to blind trust in MT outputs and the translation produced by the built-in system may by default be accepted by users as accurate and appropriate.

The popularity of testing unconventional segments therefore creates a potential risk in shaping user perceptions and experience. Successful decoding of symbolic inputs may lead users to infer that the system performs equally well when translating natural languages used in everyday communication. However, translation in real-world contexts involves far more than decoding seemingly cool symbolic representations or finding the corresponding words and phrases in another language as the purpose that a dictionary would serve. It requires interpretation of meaning within specific social, cultural, and pragmatic contexts. Treating translation primarily as a technical decoding process may hence obscure the situated and interpretive nature of cross-lingual communication, not to mention the baseline quality of MT outputs deployed on other Chinese social network services such as Wechat is not without problems (Luo and Li, 2022).

These findings underscore the importance of the ongoing effort for improving MT literacy – not only among translation students and educators, but also professional translators and other stakeholders – and for raising public awareness about the limitations and complexities of MT (Bindels et al., 2025; Krüger, 2023; Bowker, 2025). Encouraging more informed understandings of translation as a socially situated activity may therefore help users engage with MT outputs more critically.

[7] This phenomenon could reasonably be linked to the official post P04 that presented a string of symbols as the source text for MT and invited encoded texts from users to test the translation function (if true, this might indicate yet another flaw in the launch package discussed in Section 5.1).

### 5.3 Translation in a Socio-Technical System

User comments also reveal that perceptions of the translation feature extend beyond the quality of the translation output itself. Some expressed admiration for the rapid development cycle, while others voiced concern about the workload placed on developers and programmers. Such remarks indicate that users are aware of the humans involved in producing and maintaining the system, even when interacting with automated functions.

At the same time, practical concerns regarding accessibility were frequently raised. Several users reported difficulties accessing the translation feature due to version differences, missing language options, or unclear interface settings. These comments suggest that gaps in communication about feature availability or system requirements can significantly shape user experience. In this sense, perceptions of translation quality are embedded within a broader package of technical, organisational, and informational factors rather than being determined solely by linguistic accuracy.

These observations highlight the inherently social nature of translation technologies in digital environments. MT on social platforms emerges through the interaction of multiple stakeholders, including users and marketing specialists at the forefront and MT specialists and linguists behind the architecture, in a situated socio-economic context. Each group of stakeholders contributes to how translation functions are designed, implemented, interpreted, and evaluated. Consequently, in a real-life context as such, the assessments of translation quality should move beyond laboratory-style benchmarks. It is worth reiterating that while these measures remain important for comparing system performance and tracking technical improvements, they capture only part of how translation systems function in practice. This is particularly important because even though MT outputs can contain errors and impose risks across communicative contexts, its convenience can in fact 'shape perceptions and expectations of what translations should be and the different roles they can play' (Vieira and Al Sharou, 2025).

## 6 Conclusion

This study examined early user responses to the introduction of Xiaohongshu's built-in MT feature. The findings suggest that user perceptions of translation functions are shaped by a range of socio-technical factors, including promotional materials of the tools, testing practices, and feature accessibility. In this sense, translation features on social media platforms are experienced not merely as technical outputs but as components of a broader communicative ecosystem. The analysis also indicates that the types of text segments users selected to test the system were relatively limited. As a result, users' evaluations of MT performance may reflect a narrow range of use cases, potentially shaping perceptions of the function's usefulness in everyday cross-lingual communication, calling for better MT literacy among the general public, particularly in contexts where MT tools are increasingly integrated into daily interactions.

More broadly, real-world deployments of translation features generate large-scale user interactions that provide valuable insight into usability, perceived reliability, and communicative effectiveness. User experimentation, feedback, and informal testing collectively form a form of situated evaluation that reflects how translation technologies operate within everyday communication. Incorporating these perspectives into research and development may therefore contribute to more user-centred approaches to MT design and evaluation, ensuring that technological advancements are aligned with the communicative needs and expectations of the communities that use them.

Admittedly, this study provides only a snapshot of how users responded to an LLM-enabled MT feature when it was first introduced on Xiaohongshu. Further research is needed to develop a more comprehensive understanding of how the function is used in real-time communication beyond the context of user engagement with the 11 official posts analysed here. Nevertheless, it is clear that addressing the challenges of deploying MT systems on social media platforms will require closer interdisciplinary collaboration, bringing together computer scientists, translation scholars, users, and other stakeholders. Improving MT performance in practice ultimately depends not only on advances in algorithms, but also on sustained dialogue between those who build these systems and those who study and use them. As Kenny (2025) observes, 'it is not the technology alone that shapes the future; rather it is the way in which it is accommodated by the socio-cultural, legal and economic context, itself shifting in line with technological change that will have the greatest bearing'.

## A Posts Info

Below is an overview of the 11 posts by the official accounts of Xiaohongshu announcing or promoting the built-in translation functions.

- P01: 小红书翻译功能上线啦！(Xiaohongshu’s Translation Function is Online), by 日常薯 (Daily Shu), 19 January 2025.
- P02: 小红书翻译功能来啦！Translation is coming (Xiaohongshu’s Translation Function is Here), by 搜搜薯 (Search Shu), 18 January 2025.
- P03: 爆料！本队和外国朋友丝滑聊天的秘诀是 (Breaking news! My secret method to chat smoothly with foreign friends is), by 薯队长 (Captain Shu), 19 January 2025.
- P04: 请翻译：□●□○○●▲ (Please translate: □●□○○●▲), by 热点薯 (Hot topic Shu), 23 January 2025.
- P05: 小红书聊天消息也支持翻译啦！(Messages on Xiaohongshu Chats also Support Translation Now), by 日常薯 (Daily Shu), 21 January 2025.
- P06: 聊天消息翻译上线了！跨国聊天易如反掌 (Chat Message Translation is Now Online! Cross-country Chatting Becomes Effortless), by 薯队长 (Captain Shu), 22 January 2025.
- P07: 急... 外国朋友比我还懂甄嬛传！(Urgent···Foreign Friends Understand Empresses in the Palace Better Than I Do), by 娱乐薯 (Entertainment Shu), 22 January 2025.
- P08: 薯薯不语，只一味上新 (Shushu Doesn’t Say Anything, but Constantly Releasing Updates), by 发发薯 (Posting Shu), 21 January 2025.
- P09: 地球村畅行神器【双语文字配图】上线！(A Global Communication Tool [Bilingual Text-with-image Feature] Is Now Online), by 小红书创作助手 (Xiaohongshu Creator Assistant), 21 January 2025.
- P10: 全世界的创作者看过来，这个功能都会让你的创作变得更容易！(Creators Around the World, Take a Look, This Feature Will Make Your Content Creation Much Easier), by 发发薯 (Posting Shu), 20 January 2025.
- P11: 创作者有福了！本薯发现了双语字幕新功能 (Good News for Creators! I Found A New Bilingual Subtitling Feature), by 小红书创作助手 (Xiaohongshu Creator Assistant), 20 January 2025.

## B Prompt Template

You are analysing user comments from Xiaohongshu posts announcing the launch of a translation function. Your task is to classify each comment along two dimensions.

COMMENT_TYPE

- Opinion: the user expresses evaluation, reaction, or judgement about the translation function.
- Test: the user inputs words, symbols, numbers, or sentences mainly to try the translation function.
- Other: unrelated conversation, platform requests, tagging friends, emojis only, or unclear meaning.

SENTIMENT Only assign sentiment if COMMENT_TYPE = OPINION.

- Positive
- Neutral
- Negative

If COMMENT_TYPE is Test or Other, set SENTIMENT = NA.

Decision rules:

1. If the comment mainly contains random text, foreign language, symbols, numbers, or short phrases used to try translation, classify as Test.
2. If the comment evaluates or reacts to the translation function, classify as Opinion.
3. If the comment does not relate to the translation feature, classify as Other.
4. If both Test and Opinion signals appear, prioritise Opinion.

Emoji handling: Emoji-like tokens such as [点赞 R], [哇 R], [哭惹 R], [doge], [笑哭 R], [汗颜 R] often convey sentiment and should be considered when determining the sentiment of the comment.

Interpretation instruction: First briefly interpret the meaning of the comment internally (including emojis, sarcasm, or slang) before deciding the labels. Do not output the interpretation.

Output requirements: Only use the exact labels listed above. Do not invent new labels.

Return the result strictly in JSON format:

```
{
  "comment_type": "Opinion|Test|Other",
  "sentiment": "Positive|Neutral|Negative|NA"
}
```

Examples:

- Comment: 哇塞！有了！
  Output: `{"comment_type":"Opinion", "sentiment":"Positive"}`
- Comment: 不多说，让老板给你们加钱 [doge]
  Output: `{"comment_type":"Opinion", "sentiment":"Positive"}`
- Comment: 但是有的带了表情包就看不了
  Output: `{"comment_type":"Opinion", "sentiment":"Neutral"}`
- Comment: 小红书怎么被翻译成了红笔记 [笑哭 R]
  Output: `{"comment_type":"Opinion", "sentiment":"Negative"}`
- Comment: Le français peut-il aussi être traduit ?
  Output: `{"comment_type":"Test", "sentiment":"NA"}`
- Comment: 我和氧怦的好友标 [鄙视 R]
  Output: `{"comment_type":"Other", "sentiment":"NA"}`

Now classify the following comment: `{content}`